\pdfoutput=1
\documentclass[11pt]{article}

\usepackage[margin=1in]{geometry}
\usepackage{hyperref}
\usepackage{url}
\usepackage{cite}
\usepackage{amsmath,amssymb}
\usepackage{booktabs}
\usepackage{caption}
\usepackage{subcaption}
\usepackage{listings}
\usepackage{xcolor}
\usepackage{microtype}
\usepackage{tikz}
\usetikzlibrary{shapes.geometric, arrows.meta, positioning, calc, fit, backgrounds, matrix, shadows}

\title{\textbf{Proving the Utility of Large Language Models in Cybersecurity Simulations:\\A Comprehensive Examination}}
\author{
  \textbf{Stylianos Kampakis}\textsuperscript{1}, 
  \textbf{Fabio Rovai}\textsuperscript{1}, 
  \textbf{Marcos Charalambides}\textsuperscript{2},\\
  \textbf{Theodosis Mourouzis}\textsuperscript{2}, 
  \textbf{Chris Hicks}\textsuperscript{3}\\[0.5em]
  \textsuperscript{1}The Tesseract Academy, London, UK\\
  \texttt{\{stelios, fabio\}@thetesseractacademy.com}\\[0.3em]
  \textsuperscript{2}Electi Consulting, Nicosia, Cyprus\\
  \texttt{\{marcos.charalambides, theodosis.mourouzis\}@electiconsulting.com}\\[0.3em]
  \textsuperscript{3}The Alan Turing Institute, London, UK\\
  \texttt{c.hicks@turing.ac.uk}
}

\date{\today}

\begin{document}

\maketitle

\begin{abstract}
Cyber threats continue to escalate in both frequency and sophistication, necessitating more adaptive and scalable defense strategies. This paper explores how Large Language Models (LLMs) can bolster cybersecurity simulations by automating the creation of synthetic environments and identifying latent vulnerabilities. We employ YAML as a structured representation format for simulating complex network configurations, thereby enabling Large Language Model-driven pipelines to support and improve reinforcement learning (RL) agent training. Comparative studies examine the advantages of LLM-based techniques over classical approaches such as Double Q-learning with Prioritized Experience Replay (PER), emphasizing increased efficiency, higher adaptability, and enhanced realism in cyberattack simulations. In empirical benchmarks across multiple synthetic topologies, LLM-instantiated Python agents achieved up to a 94.5\% compromise rate while executing in 0.02--0.06 seconds per assessment---a $\sim$25,000$\times$ to 50,000$\times$ speedup over traditional RL training cycles. Our findings underscore the transformative potential of integrating LLMs into cybersecurity research, ultimately paving the way for more intelligent and robust cyber-defense systems.
\end{abstract}

\section{Introduction}
The cybersecurity landscape has undergone a paradigm shift, wherein threat actors leverage increasingly sophisticated tactics to exploit vulnerabilities in both conventional and emerging technological infrastructures. Traditional defense mechanisms, largely reliant on static rule sets and signature-based methods, struggle to adapt against zero-day exploits and polymorphic threats \cite{handa2019machine,dasgupta2022machine}. Consequently, there is an urgent need for advanced simulation frameworks that can accurately represent heterogeneous network environments, facilitate continuous adaptation, and support the rapid development of novel defensive strategies.

Cybersecurity simulations offer controlled, replicable environments for research and development, enabling experimentation with attack vectors, defense algorithms, and training methodologies. Among these approaches, reinforcement learning (RL) has gained prominence, particularly through Deep Reinforcement Learning (DRL) methods \cite{nguyen2021deep,limmen_awesome_rl_cybersecurity}. However, a major bottleneck lies in the generation of realistic and scalable simulation environments; manual configuration often proves time-consuming, error-prone, and limited in scope.

Recent advances in Large Language Models (LLMs) hold promise for addressing this bottleneck. Contemporary LLMs demonstrate a remarkable capacity for contextual understanding, structured data generation, and direct code synthesis \cite{ma2018language,chang2017sceneseer,long2023can,jiralerspong2024efficient}. Nevertheless, minimal prior work has explored the application of LLMs to automate the generation of detailed cybersecurity environments. To bridge this gap, the present study proposes an LLM-driven pipeline that uses YAML as a flexible and human-readable configuration scheme, thereby empowering diverse RL agents to train in dynamically generated attack-defense scenarios.

\subsection{Challenges with Existing Methods}
\begin{enumerate}
    \item \textbf{Manual Labor Intensity:} Building sophisticated simulation environments with multiple subnets, firewalls, and host vulnerabilities is highly resource-intensive. Many existing approaches rely on skilled domain experts, which becomes infeasible for larger-scale or frequently updated simulations.
    \item \textbf{Limited Adaptability:} As networks evolve (e.g., expansions, reconfigurations, or introduction of novel technologies), existing digital twins must be manually revised to maintain fidelity, incurring continuous overhead \cite{wu2021digital}.
    \item \textbf{Computational and Maintenance Costs:} Designing high-fidelity simulation frameworks necessitates extensive computational resources \cite{segovia2022design}, hindering the routine testing and training of RL algorithms.
\end{enumerate}

\subsection{Proposed Approach and Contributions}
This paper posits that Large Language Models (LLMs) can alleviate these constraints by automatically generating YAML-based cybersecurity environments. Specifically, we present:
\begin{enumerate}
    \item \textbf{A Novel LLM-Driven Pipeline:} We detail an approach that leverages prompt engineering, ``golden'' reference examples, and iterative topological validation to create realistic network topologies, vulnerability distributions, and security policies.
    \item \textbf{Comparative Analysis with Classical Methods:} We benchmark LLM-generated attack agents against conventional Double Q-learning with Prioritized Experience Replay across 6 distinct network topologies (6,000 total evaluation episodes), quantifying improvements in compromise rate, adaptability, and computational efficiency.
    \item \textbf{Insight into In-Context Learning and Robustness:} We document key empirical lessons from prompt design---including the ``memorize after debugging'' pitfall---and propose architectural enhancements such as retrieval-augmented generation (RAG).
\end{enumerate}

\section{Related Work}\label{sec:related_work}
Research in cybersecurity simulations has been influenced by multiple paradigms, including machine learning-based intrusion detection systems, digital twins for replicating physical systems, and multi-agent frameworks for coordinating attack and defense strategies.

\subsection{Machine Learning in Cybersecurity}
Machine learning approaches have significantly advanced the identification and mitigation of cyber threats, spanning intrusion detection, malware analysis, and anomaly detection \cite{handa2019machine,dasgupta2022machine}. Deep Reinforcement Learning methods have further demonstrated efficacy in adapting to dynamic adversarial patterns by updating defense strategies in response to novel attack behaviors \cite{nguyen2021deep}.

\subsection{Digital Twins and Simulation Environments}
The concept of a digital twin, a virtual replica of physical or cyber-physical systems, has been applied in cybersecurity to enable environment virtualization \cite{faleiro2022digital,eckhart2018towards}. While frameworks such as CyberBattleSim \cite{microsoft2021cyberbattlesim}, YAWNING-TITAN \cite{yawningtitan2024}, NASim \cite{schwartz2024nas}, PrimAITE \cite{arcd2023primaite}, CSLE \cite{hammar2023csle}, and NetSecGame \cite{rigaki2023out} offer modular simulation settings, they still require manual or semi-automated processes for scenario generation. This limitation narrows the scope of experimentation and hinders rapid iteration. Additionally, scaling such simulations remains non-trivial due to reconfiguration demands and alignment with real-world network properties.

\subsection{LLMs in Cybersecurity}
Recent literature explores using LLMs for specialized tasks, including automated exploit generation, detection of zero-day vulnerabilities, and contextual intrusion detection \cite{rigaki2023out,tann2023using,ferrag2023revolutionizing,purba2023software,yamin2024applications}. However, few studies have harnessed LLMs for environment creation. The success of LLMs in generating structured data in other domains (e.g., 3D scene generation \cite{ma2018language}, supply chain optimization \cite{li2023large}, and reward design \cite{ma2023eureka}) motivates our investigation into automated cybersecurity environment generation.

\subsection{Structured Output Generation with LLMs}
Generating accurate, well-formed YAML files exemplifies a structured output task. Several approaches to structured generation exist:
\begin{enumerate}
    \item \textit{Model Fine-Tuning}: Continually train a foundation model on domain-specific, example-based data to produce more consistent structured outputs \cite{zhang2024scaling,jeong2024fine,wei2021finetuned}.
    \item \textit{Grammar-Constrained Decoding}: Leverage domain-specific languages (DSLs) and grammar-enforcing token generation \cite{timlrx_generating_structured_output,microsoft2023guidance,srilab2023lmql,dottxt2024outlines,ggerganov2023llamacpp}.
    \item \textit{Schema Engineering}: Use type or interface definitions to guide and validate generation outputs \cite{microsoft2023typechat}.
\end{enumerate}
We adapt a hybrid approach that uses prompt engineering and example-based configuration to encourage syntactically valid and semantically coherent YAML generation, validated by domain-specific checks.

\section{Methodology}\label{sec:methodology}
In this section, we detail our system design for automated cybersecurity environment generation and the subsequent integration of reinforcement learning (RL) agents. We also describe the metrics used to evaluate the quality of generated environments and agent performance.

\subsection{System Architecture}
\begin{figure}[htbp]
\centering
\begin{tikzpicture}[
    scale=0.9, every node/.style={transform shape},
    node distance=0.9cm and 1.2cm,
    block/.style={rectangle, draw=blue!80!black, fill=blue!6, rounded corners=4pt, thick, text width=3.0cm, align=center, minimum height=1.1cm, font=\footnotesize},
    failblock/.style={rectangle, draw=red!75!black, fill=red!6, rounded corners=4pt, thick, text width=3.0cm, align=center, minimum height=1.1cm, font=\footnotesize},
    passblock/.style={rectangle, draw=green!60!black, fill=green!7, rounded corners=4pt, thick, text width=3.0cm, align=center, minimum height=1.1cm, font=\footnotesize},
    arrow/.style={-{Latex[length=2.5mm]}, thick, draw=gray!70!black}
]
\node[block] (p1) {\textbf{Approach 1 (Template)}\\Generic schema prompt with placeholders};
\node[block, right=of p1] (llm1) {\textbf{ChatGPT-4}\\\textit{Zero-shot generation}};
\node[block, right=of llm1] (nasim1) {\textbf{NASim Validator}\\\textit{Schema \& topology check}};
\node[failblock, right=of nasim1] (out1) {\textbf{Frequent Failures}\\Adjacency \& firewall mismatch (40\% valid)};

\node[block, below=1.2cm of p1] (p2) {\textbf{Approach 2 (Example)}\\Few-shot prompt with golden YAML};
\node[block, right=of p2] (llm2) {\textbf{ChatGPT-4}\\\textit{Context-guided mutation}};
\node[block, right=of llm2] (nasim2) {\textbf{NASim Validator}\\\textit{Topology reinforcement}};
\node[passblock, right=of nasim2] (out2) {\textbf{High Fidelity}\\Consistent relational bindings (70\% valid)};

\draw[arrow] (p1) -- (llm1);
\draw[arrow] (llm1) -- (nasim1);
\draw[arrow] (nasim1) -- (out1);

\draw[arrow] (p2) -- (llm2);
\draw[arrow] (llm2) -- (nasim2);
\draw[arrow] (nasim2) -- (out2);
\end{tikzpicture}
\caption{High-level architecture for YAML file creation and experimentation. This diagram contrasts Approach~1 (Template-Based Generation) with Approach~2 (Example-Based Generation).}
\label{fig:high_level_arch}
\end{figure}
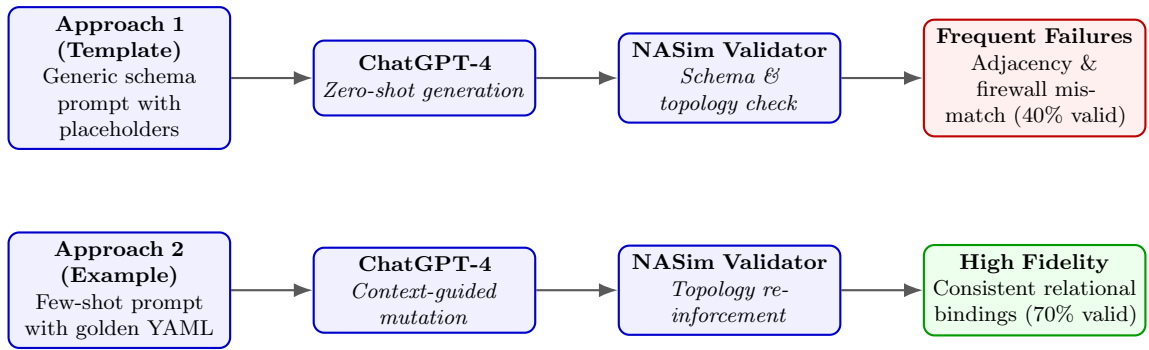

We develop a pipeline that automatically generates YAML configurations using an LLM and tests them within a cybersecurity emulator (e.g., NASim \cite{schwartz2024nas}). Figure~\ref{fig:high_level_arch} illustrates two main approaches for YAML generation:
\begin{enumerate}
    \item \textbf{Template-Based Configuration (Approach 1):} Uses rigid YAML schemas that the LLM fills in according to user prompts. Under this paradigm, the model is prompted with generic structural placeholders.
    \item \textbf{Example-Based Configuration (Approach 2):} Starts from one or more ``golden'' reference YAML files. The LLM edits or extends them in response to user instructions (e.g., varying host counts, subnets, and service combinations), achieving substantially higher success rates.
\end{enumerate}

\subsubsection{LLM-Driven YAML Generation}
\begin{figure}[htbp]
\centering
\begin{tikzpicture}[
    scale=0.88, every node/.style={transform shape},
    node distance=0.8cm and 1.0cm,
    stepnode/.style={rectangle, draw=blue!70!black, fill=blue!5, rounded corners=4pt, thick, text width=2.6cm, align=center, minimum height=1cm, font=\footnotesize},
    decision/.style={diamond, draw=orange!80!black, fill=orange!8, aspect=1.8, thick, align=center, font=\scriptsize, inner sep=1pt},
    passnode/.style={rectangle, draw=green!60!black, fill=green!8, rounded corners=4pt, thick, text width=2.4cm, align=center, minimum height=0.9cm, font=\footnotesize},
    failnode/.style={rectangle, draw=red!70!black, fill=red!6, rounded corners=4pt, thick, text width=2.4cm, align=center, minimum height=0.9cm, font=\footnotesize},
    arrow/.style={-{Latex[length=2.2mm]}, thick, draw=gray!80!black}
]
\node[stepnode] (in) {\textbf{Scenario Prompt}\\Requirements \& documentation};
\node[stepnode, right=of in] (llm) {\textbf{LLM Engine}\\Candidate YAML synthesis};
\node[decision, right=of llm] (dec1) {\textbf{Topology}\\ \textbf{Check}};
\node[stepnode, right=1.0cm of dec1] (emu) {\textbf{NASim Test}\\Emulator load};
\node[decision, below=0.8cm of emu] (dec2) {\textbf{Runs in}\\ \textbf{NASim?}};
\node[failnode, below=0.8cm of llm] (discard) {\textbf{DISCARDED}\\Rejected candidate};
\node[passnode, below=0.8cm of in] (stored) {\textbf{STORED}\\Golden YAML Pool};

\draw[arrow] (in) -- (llm);
\draw[arrow] (llm) -- (dec1);
\draw[arrow] (dec1) -- node[above, font=\scriptsize]{Pass} (emu);
\draw[arrow] (dec1) -- node[right, font=\scriptsize]{Fail} (discard);
\draw[arrow] (emu) -- (dec2);
\draw[arrow] (dec2) -- node[above, font=\scriptsize]{Fail} (discard);
\draw[arrow] (dec2) -| node[below, font=\scriptsize, pos=0.35]{Pass} (stored);
\end{tikzpicture}
\caption{Process flow for YAML creation, including topological reinforcement. If the configuration passes validation and successfully runs in the NASim emulator, it is stored; otherwise, it is discarded.}
\label{fig:yaml_creation}
\end{figure}
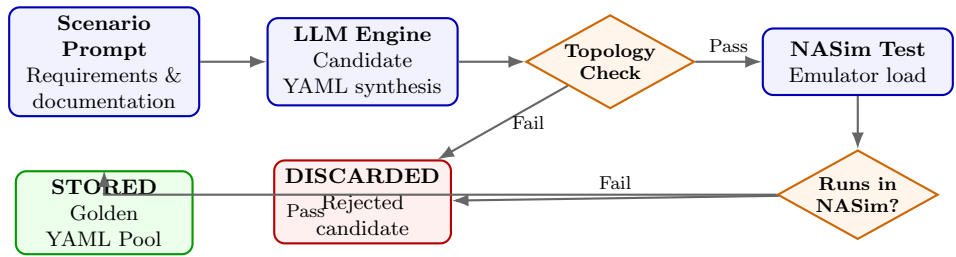

After receiving a prompt (including a documentation link or special instructions), the LLM generates a candidate YAML configuration. We apply topological reinforcement checks to ensure coherence and realism, discarding outputs that fail these checks (Figure~\ref{fig:yaml_creation}). This includes network adjacency matrix correctness, firewall bidirectional rule consistency, vulnerability definitions, and process mapping.

\subsubsection{Generated YAML Topology Specification}
Listing~\ref{lst:yaml_example} illustrates a representative synthetic network configuration generated by our LLM pipeline. The schema defines subnet partitioning, topology adjacency, high-value sensitive targets, available exploits, privilege escalation vectors, individual host profiles, and directional firewall filters.

\begin{lstlisting}[language=Python, caption={Sample LLM-Generated YAML Network Configuration (AI YAML 1).}, label={lst:yaml_example}]
subnets: [1, 1, 1, 1, 1]
topology: [[0, 1, 1, 1, 0, 0], [1, 0, 0, 0, 1, 1], [1, 0, 0, 0, 1, 1], 
           [1, 0, 0, 0, 1, 1], [0, 1, 1, 1, 0, 0], [0, 1, 1, 1, 0, 0]]
sensitive_hosts: {(2, 0): 100, (5, 0): 100}
os: [linux, windows]
services: [ssh, http]
processes: [tomcat, apache, iis]
exploits:
  e_ssh:  {service: ssh,  os: linux,   prob: 0.8, cost: 1, access: user}
  e_http: {service: http, os: windows, prob: 0.7, cost: 1, access: user}
privilege_escalation:
  pe_iis: {process: iis,  os: windows, prob: 0.9, cost: 1, access: root}
service_scan_cost: 1
os_scan_cost: 1
subnet_scan_cost: 1
process_scan_cost: 1
host_configurations:
  (1, 0): {os: linux,   services: [ssh, http], processes: [tomcat, apache], 
           firewall: {(4, 0): [ssh], (5, 0): [http]}}
  (2, 0): {os: windows, services: [http, ssh], processes: [apache, iis], 
           firewall: {(4, 0): [http], (5, 0): [ssh]}}
  (3, 0): {os: linux,   services: [ssh],       processes: [tomcat, apache], 
           firewall: {(4, 0): [ssh], (5, 0): [http]}}
  (4, 0): {os: windows, services: [http],      processes: [iis], 
           firewall: {(1, 0): [http], (2, 0): [ssh]}}
  (5, 0): {os: linux,   services: [ssh],       processes: [tomcat, iis], 
           firewall: {(1, 0): [http], (2, 0): [ssh]}}
firewall: {(0, 1): [ssh], (1, 0): [], (0, 2): [http], (2, 0): [], 
           (0, 3): [http], (3, 0): [], (1, 4): [ssh, http], (4, 1): [], 
           (1, 5): [http], (5, 1): [ssh], (4, 5): [ssh, http], (5, 4): []}
step_limit: 1000
\end{lstlisting}

\subsection{Reinforcement Learning Integration}
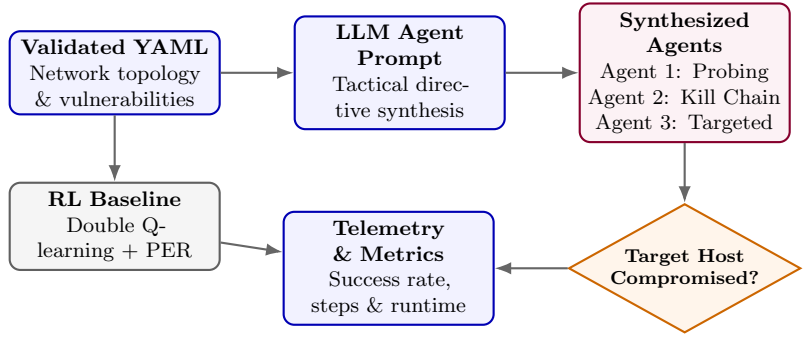
\begin{figure}[htbp]
\centering
\begin{tikzpicture}[
    scale=0.88, every node/.style={transform shape},
    node distance=0.8cm and 1.1cm,
    stepnode/.style={rectangle, draw=blue!70!black, fill=blue!5, rounded corners=4pt, thick, text width=2.9cm, align=center, minimum height=1cm, font=\footnotesize},
    decision/.style={diamond, draw=orange!80!black, fill=orange!8, aspect=1.8, thick, align=center, font=\scriptsize, inner sep=1pt},
    agentnode/.style={rectangle, draw=purple!70!black, fill=purple!5, rounded corners=4pt, thick, text width=2.9cm, align=center, minimum height=1cm, font=\footnotesize},
    bench/.style={rectangle, draw=gray!75!black, fill=gray!8, rounded corners=4pt, thick, text width=2.9cm, align=center, minimum height=1cm, font=\footnotesize},
    arrow/.style={-{Latex[length=2.2mm]}, thick, draw=gray!80!black}
]
\node[stepnode] (spec) {\textbf{Validated YAML}\\Network topology \& vulnerabilities};
\node[stepnode, right=of spec] (llm) {\textbf{LLM Agent Prompt}\\Tactical directive synthesis};
\node[agentnode, right=of llm] (agents) {\textbf{Synthesized Agents}\\Agent 1: Probing\\Agent 2: Kill Chain\\Agent 3: Targeted};
\node[bench, below=1.0cm of spec] (rl) {\textbf{RL Baseline}\\Double Q-learning + PER};
\node[decision, below=0.9cm of agents] (eval) {\textbf{Target Host}\\ \textbf{Compromised?}};
\node[stepnode, left=of eval] (telemetry) {\textbf{Telemetry \& Metrics}\\Success rate, steps \& runtime};

\draw[arrow] (spec) -- (llm);
\draw[arrow] (llm) -- (agents);
\draw[arrow] (agents) -- (eval);
\draw[arrow] (eval) -- (telemetry);
\draw[arrow] (rl) -- (telemetry);
\draw[arrow] (spec) -- (rl);
\end{tikzpicture}
\caption{Agent creation and evaluation process. Once a valid YAML configuration is verified, adversarial agents are synthesized and deployed to emulate attack campaigns against classical RL baselines.}
\label{fig:agent_creation}
\end{figure}

After generating a valid YAML file, we instantiate RL-based agents and LLM-synthesized attack scripts to attempt to compromise the environment. Figure~\ref{fig:agent_creation} shows how these agents are constructed and benchmarked against standard reinforcement learning baselines.

\subsection{Overall System Pipeline}
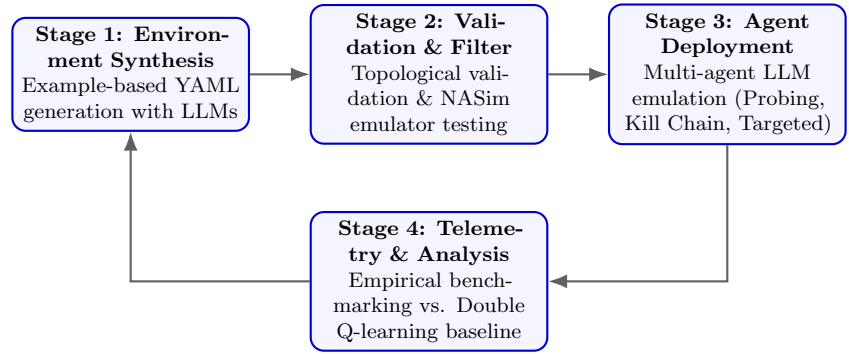
\begin{figure}[htbp]
\centering
\begin{tikzpicture}[
    scale=0.88, every node/.style={transform shape},
    node distance=0.7cm and 0.9cm,
    stage/.style={rectangle, draw=blue!80!black, fill=blue!4, rounded corners=5pt, thick, text width=3.3cm, align=center, minimum height=1.3cm, font=\footnotesize},
    arrow/.style={-{Latex[length=2.5mm]}, thick, draw=gray!75!black}
]
\node[stage] (s1) {\textbf{Stage 1: Environment Synthesis}\\Example-based YAML generation with LLMs};
\node[stage, right=of s1] (s2) {\textbf{Stage 2: Validation \& Filter}\\Topological validation \& NASim emulator testing};
\node[stage, right=of s2] (s3) {\textbf{Stage 3: Agent Deployment}\\Multi-agent LLM emulation (Probing, Kill Chain, Targeted)};
\node[stage, below=1.0cm of s2] (s4) {\textbf{Stage 4: Telemetry \& Analysis}\\Empirical benchmarking vs. Double Q-learning baseline};

\draw[arrow] (s1) -- (s2);
\draw[arrow] (s2) -- (s3);
\draw[arrow] (s3) |- (s4);
\draw[arrow] (s4) -| (s1);
\end{tikzpicture}
\caption{End-to-end closed-loop workflow of the system, from prompt engineering and LLM-based YAML generation to agent emulation, telemetry collection, and iterative refinement.}
\label{fig:complete_architecture}
\end{figure}

Figure~\ref{fig:complete_architecture} synthesizes the entire workflow. By integrating YAML generation and agent testing into a single closed loop, we can rapidly iterate over multiple scenarios, store successful outputs, and feed back any errors into subsequent prompt refinements.

\subsection{Agent Framework}
We develop three specialized attack agents prompted under the scenario prompt:
\begin{quote}
    \emph{“I am working with configuration network files in Python. I would like you to develop an advanced approach to emulate this environment and identify potential vulnerabilities or cracks in the shortest possible time. You can use any techniques, such as reinforcement learning, brute force, or other state-of-the-art methods. Write in many steps how the network was compromised or if it was not compromised. Write `this network was compromised’ or `not compromised.’ Please provide python code only. Read following configurations without rewriting it.”}
\end{quote}

The resulting agents express three distinct tactical behaviors:
\begin{enumerate}
    \item \textbf{Agent 1 (Step-by-Step Probing):} Methodical breadth-first scanning, enumeration of services, operating systems, and active processes across subnets before attempting exploitation. Performs exhaustive exploration to ensure all reachable attack surfaces are mapped. Incorporates probabilistic execution checks reflecting real-world exploit success uncertainty.
    \item \textbf{Agent 2 (Cyber Kill Chain \& Network Propagation):} Simulates multi-stage attack propagation (reconnaissance $\rightarrow$ weaponization $\rightarrow$ exploit $\rightarrow$ lateral movement $\rightarrow$ C2) starting from pre-compromised beachhead nodes. Explicitly evaluates network segmentation and subnet-to-subnet firewall filtering rules to determine viable lateral movement paths.
    \item \textbf{Agent 3 (Targeted Exploit \& Sensitive Host Focus):} Direct exploit execution targeting known high-impact services without exhaustive host-by-host scanning. Minimizes observable network footprints by prioritizing rapid privilege escalation paths directly to defined sensitive target hosts, terminating simulation immediately upon target breach.
\end{enumerate}

\subsection{Comparative Baseline: Double Q-Learning with PER}
To establish an empirical baseline, we implemented a classical reinforcement learning agent utilizing \textbf{Double Q-learning with Prioritized Experience Replay (PER)} \cite{vanhasselt2016deep,schaul2015prioritized} interfacing directly with the NASim environment:
\begin{itemize}
    \item \textbf{Double Q-Learning:} To eliminate the overestimation bias inherent in standard Q-learning, the agent maintains two independent Q-value estimators, $Q_1(s, a)$ and $Q_2(s, a)$. During each update, one table selects the greedy action while the alternate table estimates its value:
    \begin{equation}
        Y_t^{DoubleQ} = R_{t+1} + \gamma Q_2\left(S_{t+1}, \arg\max_{a} Q_1(S_{t+1}, a)\right)
    \end{equation}
    \item \textbf{Prioritized Experience Replay (PER):} Experiences $e_t = (s_t, a_t, r_t, s_{t+1}, d_t)$ are stored in a prioritized buffer where transition sampling probabilities $P(i)$ are proportional to temporal-difference (TD) error magnitudes:
    \begin{equation}
        P(i) = \frac{p_i^\alpha}{\sum_k p_k^\alpha}, \quad p_i = |\delta_i| + \epsilon_{per}
    \end{equation}
    \item \textbf{Exploration Schedule:} The agent employs an $\epsilon$-greedy exploration policy, initialized at $\epsilon_0 = 1.0$ and decaying dynamically over training episodes towards $\epsilon_{min} = 0.05$.
    \item \textbf{Training and Logging:} Policy convergence, episode returns, and TD errors were logged in real-time using TensorBoard via \texttt{SummaryWriter}.
\end{itemize}

\section{Results and Findings}\label{sec:results}

\subsection{YAML Generation Success}
Table~\ref{tab:yaml_generation} summarizes the validation success rates across 25 independent generation trials per method. The \textit{example-based approach} achieved a \textbf{70\% success rate} in producing valid configurations without manual debugging, outperforming prompt-only and template-only strategies that suffered from recurring firewall rule mismatches and adjacency matrix inconsistencies.

\begin{table}[htbp]
\centering
\small
\caption{LLM-Generated YAML Success Rates (25 runs per method).}
\label{tab:yaml_generation}
\begin{tabular}{lcc}
\toprule
\textbf{Approach} & \textbf{Valid Configs (\%)} & \textbf{Common Failure Mode}\\
\midrule
Template-Based (Approach 1) & 40\% & Syntax errors and missing keys \\
Prompt-Based                & 55\% & Firewall rule mismatches \\
Example-Based (Approach 2)  & \textbf{70\%} & Occasional incomplete adjacency \\
\bottomrule
\end{tabular}
\end{table}

\subsection{Comprehensive Scenario-by-Scenario Benchmark}
We conducted an extensive empirical benchmark across 6 distinct network environments: the standard NASim Tiny Example benchmark and 5 AI-generated synthetic topologies (AI YAML 1 through 5). Each scenario was evaluated over 1,000 independent attack episodes per agent (6,000 evaluation episodes total). Table~\ref{tab:scenario_benchmark} details the full breakdown of successful breaches, failures, and execution times.

\begin{table}[htbp]
\centering
\footnotesize
\setlength{\tabcolsep}{3pt}
\caption{Comprehensive Performance Benchmark Across 6 Network Scenarios (1,000 runs per agent per scenario; 6,000 runs total).}
\label{tab:scenario_benchmark}
\begin{tabular}{llcccc}
\toprule
\textbf{Scenario} & \textbf{Approach} & \textbf{Unsuccessful} & \textbf{Successful} & \textbf{Success (\%)} & \textbf{Time (s)}\\
\midrule
\textbf{Tiny Example}   & Double Q-Learning & 0   & 1    & 100.0\% & 914.11 \\
(NASim Benchmark)       & Agent 1 (Probing) & 48  & 952  & 95.2\%  & \textbf{0.03} \\
                        & Agent 2 (Kill Chain) & 0 & 1000 & \textbf{100.0\%} & \textbf{0.03} \\
                        & Agent 3 (Targeted)   & 50 & 950  & 95.0\%  & \textbf{0.02} \\
\midrule
\textbf{vs 1st AI Gen YAML} & Double Q-Learning & 1   & 0    & 0.0\%   & 1452.38 \\
                            & Agent 1 (Probing) & 381 & 619  & 61.9\%  & \textbf{0.06} \\
                            & Agent 2 (Kill Chain) & 0 & 1000 & \textbf{100.0\%} & \textbf{0.05} \\
                            & Agent 3 (Targeted)   & 348 & 652  & 65.2\%  & \textbf{0.04} \\
\midrule
\textbf{vs 2nd AI Gen YAML} & Double Q-Learning & 1   & 0    & 0.0\%   & 1716.36 \\
                            & Agent 1 (Probing) & 387 & 613  & 61.3\%  & \textbf{0.06} \\
                            & Agent 2 (Kill Chain) & 333 & 667 & 66.7\%  & \textbf{0.06} \\
                            & Agent 3 (Targeted)   & 418 & 582  & 58.2\%  & \textbf{0.05} \\
\midrule
\textbf{vs 3rd AI Gen YAML} & Double Q-Learning & 0   & 1    & 100.0\% & 1485.40 \\
                            & Agent 1 (Probing) & 134 & 866  & 86.6\%  & \textbf{0.05} \\
                            & Agent 2 (Kill Chain) & 0 & 1000 & \textbf{100.0\%} & \textbf{0.04} \\
                            & Agent 3 (Targeted)   & 127 & 873  & 87.3\%  & \textbf{0.04} \\
\midrule
\textbf{vs 4th AI Gen YAML} & Double Q-Learning & 1   & 0    & 0.0\%   & 1477.04 \\
                            & Agent 1 (Probing) & 131 & 869  & 86.9\%  & \textbf{0.05} \\
                            & Agent 2 (Kill Chain) & 0 & 1000 & \textbf{100.0\%} & \textbf{0.05} \\
                            & Agent 3 (Targeted)   & 105 & 895  & 89.5\%  & \textbf{0.04} \\
\midrule
\textbf{vs 5th AI Gen YAML} & Double Q-Learning & 1   & 0    & 0.0\%   & 1475.28 \\
                            & Agent 1 (Probing) & 116 & 884  & 88.4\%  & \textbf{0.05} \\
                            & Agent 2 (Kill Chain) & 0 & 1000 & \textbf{100.0\%} & \textbf{0.05} \\
                            & Agent 3 (Targeted)   & 98  & 902  & 90.2\%  & \textbf{0.04} \\
\midrule
\textbf{Aggregate (All Scenarios)} & \textbf{Agent 1 (Probing)}    & \textbf{1,197} & \textbf{4,803} & \textbf{80.0\%} & \textbf{0.05 avg} \\
                                   & \textbf{Agent 2 (Kill Chain)} & \textbf{333}   & \textbf{5,667} & \textbf{94.5\%} & \textbf{0.05 avg} \\
                                   & \textbf{Agent 3 (Targeted)}   & \textbf{1,146} & \textbf{4,854} & \textbf{80.9\%} & \textbf{0.04 avg} \\
\bottomrule
\end{tabular}
\end{table}

\subsection{Agent Performance and Tactical Comparison}
Across all 6,000 test runs, the three agent strategies exhibited distinct operational characteristics:
\begin{itemize}
    \item \textbf{Agent 2 (Cyber Kill Chain)} demonstrated the highest overall robustness, achieving a \textbf{94.5\% aggregate compromise rate} and achieving 100\% compromise in 5 of the 6 topologies. Its ability to account for firewall segmentation rules allowed it to consistently find viable multi-hop pivoting routes.
    \item \textbf{Agent 1 (Probing)} achieved an \textbf{80.0\% aggregate compromise rate}. While effective in smaller topologies, its exhaustive enumeration approach resulted in occasional failures in segmented architectures where blind probing triggered firewall blocks.
    \item \textbf{Agent 3 (Targeted)} achieved an \textbf{80.9\% aggregate compromise rate} with the lowest step counts and fastest execution time (0.02--0.05 seconds). In scenarios where direct exploit paths existed, Agent~3 reached target hosts with minimal overhead.
\end{itemize}

\subsection{Execution Efficiency: LLM Synthesis vs. Classical RL}
A central finding of our benchmark is the dramatic efficiency advantage of LLM-instantiated analysis:
\begin{itemize}
    \item \textbf{Training Latency of Classical RL:} The Double Q-learning agent required between \textbf{914.11 and 1,716.36 seconds ($\sim$15 to 28.6 minutes)} per environment to train and converge policy values over thousands of exploratory steps. In 4 of the 5 AI-generated environments, the classical RL agent failed to discover optimal compromise paths within standard step limits due to sparse reward bottlenecks.
    \item \textbf{Rapid Assessment Speedup:} In contrast, LLM-generated Python agents executed complete vulnerability emulations in \textbf{0.02 to 0.06 seconds}. This represents an assessment speedup factor of approximately $\mathbf{25,000\times}$ \textbf{to} $\mathbf{50,000\times}$, demonstrating that LLMs can rapidly evaluate synthetic digital twins without requiring compute-intensive exploration phases.
\end{itemize}

\section{Discussion}\label{sec:discussion}
Our findings confirm that Large Language Models can reliably generate valid, topologically sound cybersecurity simulation environments when guided by example-based reference configurations. 

\subsection{Prompt Engineering Insights: The ``Memorize After Debugging'' Pitfall}
During our iterative prompt optimization experiments, we uncovered an important nuance in multi-turn LLM interaction:
\begin{quote}
    \emph{When the model encountered an assertion failure (such as an asymmetrical firewall rule or adjacency dimension mismatch) and was instructed to correct the error and ``memorize the corrected configuration as a good example,'' subsequent generations became noticeably more prone to subtle edge-case bugs.}
\end{quote}
This phenomenon occurs because conversational in-context correction introduces noisy conversational artifacts and contradictory structural priors into the model's active context window. Consequently, we found that maintaining a static, validated ``golden'' reference example and using strict automated rejection filtering yielded far more consistent generation fidelity than multi-turn conversational repair.

\subsection{Topological Constraints and Firewall Symmetry}
The primary source of generation failures in LLM-synthesized environments stems from relational dependencies. In NASim, every subnet connection in the adjacency matrix must be mirrored by exact bidirectional entries in the firewall dictionary. Pure zero-shot prompting frequently generated single-directional firewall rules, triggering assertion errors during emulator initialization. The example-based approach mitigates this by providing structural relational anchors that preserve matrix-dictionary alignment.

\subsection{Limitations and Generalization}
Despite high compromise rates, current evaluations primarily verify syntactic execution and reachability within emulated state spaces. In real-world enterprise infrastructures, dynamic state elements (e.g., rate limiting, intrusion detection systems, host-based firewalls) introduce stochastic noise. Future work will investigate integrating Retrieval-Augmented Generation (RAG) backed by real-world CVE databases to ensure generated vulnerabilities reflect production software vulnerabilities.

\section{Conclusion and Future Directions}\label{sec:conclusion}
This paper presented an empirical investigation into the utility of Large Language Models for automated cybersecurity simulation generation and adversarial evaluation. Our findings show that:
\begin{enumerate}
    \item \textbf{Automated Environment Generation:} LLMs can generate simulation-ready network configurations in YAML format with a 70\% out-of-the-box validity rate using example-based prompt structures.
    \item \textbf{High-Fidelity Adversarial Emulation:} LLM-prompted attack agents achieved high compromise rates (up to 94.5\%) while executing orders of magnitude faster than classical reinforcement learning algorithms.
    \item \textbf{Significant Efficiency Gains:} LLM-driven security emulation delivers a $\sim$25,000$\times$ to 50,000$\times$ speedup over traditional Double Q-learning training loops, making it an ideal approach for high-throughput digital twin vulnerability testing.
\end{enumerate}

Future research will extend this pipeline to multi-agent cyber ranges where LLM-driven red teams and blue teams adapt simultaneously in co-evolving defensive scenarios.

\section*{Acknowledgments}
We gratefully acknowledge the Alan Turing Institute and the Department for AI for CyberDefense for providing the financial support that made this research possible. We also extend our gratitude to our colleagues and domain experts whose valuable insights and feedback significantly enhanced the design and analysis of this work.

\end{document}